# Twist Angle-Dependent Bands and Valley Inversion in 2D Materials/hBN Heterostructures

Shi Che[1,2*], Petr Stepanov[1,2*], Supeng Ge[3*], Yongjin Lee[2], Kevin Myhro[2], Yanmeng Shi[2], Ruoyu Chen[1], Ziqi Pi[2], Cheng Pan[2], Bin Cheng[2], Takashi Taniguchi[4], Kenji Watanabe[4], Marc Bockrath[1,2], Yafis Barlas[2,3], Roger Lake[3], Chun Ning Lau[1,2†]

*These authors contribute equally to this work.

[1] Department of Physics, The Ohio State University, Columbus, OH 43221
[2] Department of Physics and Astronomy, University of California, Riverside, CA 92521
[3] Department of Electrical and Computer Engineering, University of California, Riverside, CA 92521
[4] National Institute for Materials Science, 1-1 Namiki Tsukuba Ibaraki 305-0044 Japan.

## ABSTRACT

The use of relative twist angle between adjacent atomic layers in a van der Waals heterostructure, has emerged as a new degree of freedom to tune electronic and optoelectronic properties of devices based on 2D materials. Using ABA-stacked trilayer (TLG) graphene as the model system, we show that, contrary to conventional wisdom, the band structures of 2D materials are systematically tunable depending on their relative alignment angle between hexagonal BN (hBN), even at very large twist angles. Moreover, addition or removal of the hBN substrate results in an inversion of the K and K' valley in TLG's lowest Landau level (LL). Our work illustrates the critical role played by substrates in van der Waals heterostructures and opens the door towards band structure modification and valley control via substrate and twist angle engineering.

---

[†] Email: lau.232@osu.edu

The recent advent of van der Waals heterostructures has created an opening for exploring and engineering an almost limitless number of low-dimensional crystals whose properties can be tailored by quantum confinement and proximity effect[1]. In addition to traditional degrees of freedom such as spin, valley and layer, a relative twist between adjacent layers, which breaks the translational symmetry of the heterostructure, provides yet another "knob" to tune the electronic and optoelectronic properties[2-6]. For instance, depending on the relative twist angles between the two adjacent layers, transitions between commensurate and incommensurate stacking at graphene/graphite interfaces have been observed to give rise to modulations in interlayer conductivity[7] and contact resistance[8].

Despite the increasing interest in the twist angle-dependent properties, experimental and theoretical exploration of the so-called "twistronics" have largely been limited to layers composed of the same material[6-9]. The only exception is that the formation of a moiré superlattice when graphene and hexagonal BN (hBN) lattices are aligned within 1º, giving rise to secondary Dirac points and the celebrated Hofstadter butterfly spectrum[10-17]. Nevertheless, at large twist angles, hBN, which is the substrate of choice for high mobility vdW devices, is typically treated as an inert substrate that exerts no influence on the supported material.

Here we show that, contrary to the conventional wisdom, the electronic band structure of a 2D material can be systematically tuned via the relative twist angle with hBN, even at very large twist angles approaching 30º. In fact, the mere addition and removal of the hBN substrates results in an inversion of the K and K' valleys in the lowest Landau level. These results are not only crucial for the proper interpretation of electronic and optoelectronic studies of 2D materials, but also demonstrates the tunability of their band structures and open the door for band structure modification via twist angle and substrate engineering.

To illustrate the effect of twist angles on the band structures, we choose ABA-stacked trilayer graphene (TLG)[18-27] as the platform. It consists of only three atomic layers but contains the entire set of the Slonczewski-Weiss-McClure (SWMc) hopping parameters for graphite[23,28,29]. Its band structure can be decomposed into the combination of a monolayer graphene (MLG)-like band and a bilayer graphene (BLG)-like band. Both bands are individually gapped, and vertically offset from each other, with MLG (BLG) band edges at $\delta-\gamma_5/2+\delta_2$ ($-2\delta_2$) and $-\gamma_2/2+\delta_2$ ($\gamma_2/2+\delta_2$), respectively, where the $\gamma$'s are the SWMc parameters, $\delta$ denotes the energy difference between stacked and unstacked atoms, and $\delta_2$ describes the potential difference between the middle layer and the average of the outer layers. In a magnetic field, the lowest Landau levels (LLs) for both bands, located at the band edges, are almost non-dispersive, while the higher $N>1$ LLs in MLG- and BLG-like bands disperse with $B^{1/2}$ and $B$, respectively, giving rise to numerous crossing points (Fig. 1a). These LL crossing points are very sensitive to and therefore used for extracting the values of the hopping parameters[30-35]. However, despite the simplicity of the system and similarity of the techniques, there are consideration variations in the extracted values[30-35], sometimes by more than an order of magnitude. Notably, similar discrepancies are found in earlier works on graphite[36,37] and more recently on bilayer graphene[38-41].

To investigate the impact of hBN and its twist angles formed with TLG, we fabricate and measure two types of high mobility devices: two-terminal suspended devices (Fig. 1b) are fabricated by coupling to Cr/Au electrodes and released from the $SiO_2$ layer by hydrofluoric acid etching[42-44], while hBN/TLG heterostructures (Fig. 1c) are fabricated by successive sacrificial layer transfers[45,46], $SF_6$ plasma etching to define Hall-bar geometry and electron beam lithography to pattern electrodes and top gates[47]. For samples supported on hBN substrates, the twist angle $\theta$ between TLG and bottom hBN is determined by the angle between the long, straight edge of the two types of atomic layers in SEM or optical images (Fig. 1d)[47]. Completed devices are measured in a $He^3$ cryostat using standard lock-in techniques at 300 mK.

Fig. 2a-c displays longitudinal resistance $R_{xx}$ as a function of charge density $n$ and magnetic field $B$ at zero out-of-plane electric field $E_\perp$ for two different hBN-supported devices (hBN1 and hBN2) and one suspended device (S1). We first examine the Landau fans of the hBN-supported samples that are fabricated and measured under nominally identical conditions. One of the devices (hBN1) exhibits secondary Dirac points at $n\sim\pm6.7\times10^{12}$ cm$^{-2}$, indicating that the presence of a Moire superlattice due to the lattice alignment between TLG flake and the hBN substrate[13-16]. Apart from this superlattice feature, both Landau fans appear *qualitatively* similar: the dark blue regions, corresponding to vanishing $R_{xx}$ and QH plateaus, fan out from the origin; superimposed on the Landau fans are a series of discrete bright yellow and blue spots that correspond to the crossing points between the LLs originated from the MLG-like and BLG-like LLs. For instance, the crossing point that is outlined by the dotted circle P in Fig. 2a arises from the crossing between the (M, 0) and the (B, 2) bands, where M and B refer to the MLG- and BLG-like branches, and the numbers indicate orbital LL index, respectively. Similarly, the crossing points Q and S correspond to the intersections of (M, 1–/+) and (B, 7–/+) on the hole and electron side, respectively.

Intriguingly, despite being nominally identical, the crossing points in these 2 devices occur at different $n$ and $B$. The most dramatic difference is the crossing point P – it occurs at the $n=3.6\times10^{11}$ cm$^{-2}$ and $B=4.1$ T for hBN1, but appears at higher $n=5\times10^{11}$ cm$^{-2}$ and $B=4.8$ T for hBN2. The details of this LL crossing can be appreciated more clearly in the zoom-in plots in Fig. 2c-d. Similarly, crossing points Q and S also differ in the 2 devices – comparing to hBN1, Q moves to lower $B$ while S to higher magnetic fields in hBN2. These crossing points that emerge at different values of $B$ in the 2 devices are indicated by pairwise vertical arrows in Fig. 2a-b.

The observed variations in LL crossing points in different devices suggest that the hopping parameters are not constants, but vary with substrates. To gain further insight into the impact of the substrate on the band structure, we examine 9 different hBN-supported devices, and plot the crossing points $B_P$, $B_Q$ and $B_S$ as a function of the twist angle $\theta$ between the TLG flake and the hBN substrate (Fig. 3a-c). Evidently, as $\theta$ decreases, *i.e.* as hBN/TLG lattices become aligned, $B_P$ and $B_S$ steadily decrease while $B_Q$ systematically increases. The movements of all 3 crossing points indicate that larger $\theta$ shifts the MLG-like bands upwards relative to the BLG-like bands, *i.e.* increases the vertical offset between these two bands, given by $\sim|\gamma_2|$. This is borne out quantitatively by our extraction of $\gamma_2$ by fitting the crossing points to calculated LL spectra, which are plotted as blue triangles in Fig. 3a (see Discussion section below for details of the calculation).

These results show that $|\gamma_2|$ increases in samples with small Moire superlattice periods, in which the graphene-hBN interaction is relatively weak. Arguably the weakest graphene-substrate interaction is achieved by removing the substrate altogether. To this end, we examine the LL fan of differentiated two-terminal conductance $dG/dB(n,B)$ for the suspended device S1 in Fig. 2e. Here the range of $n$ is limited in order to minimize the risk of collapsing the suspended membrane under large gate voltages. Nevertheless, crossing point P (indicated by the dotted circle) is visible, which appears at a much higher $B \sim 6.8$ T than that in hBN-supported devices. This large $B_P$ confirms the trend observed in hBN-supported devices, that is, as the interaction with the substrate weakens, the MLG-like bands moves to higher energies.

In addition to the crossing points, another salient difference between S1 and hBN1 emerges as the QH state at filling factor $\nu=-2$. Except at very small $B$, its gap $\Delta_{\nu=-2} \approx |\gamma_2/2 + 3\delta_2|$ arises from the *valley* gap of the lowest LLs in the BLG-like branch, *i.e.* the energetic separation between (B, K, 0/1) and (B, K', 0/1) levels. Strikingly, this QH state is extremely robust in S1 and resolved with quantized conductance at $B$ as small as 0.2 T (Fig. 2e). In contrast, in hBN-supported devices, it remains unresolved even at 8 T (Fig. 2c-d). To ensure the different $\nu=-2$ stability does not arise from sample-to-sample variations in disorder or contact resistance, we examine 8 suspended and 9 hBN-supported devices. The results are summarized in Fig. 3d, which plots $B_{min}$, the minimum $B$ at which the $\nu=-2$ state is resolved, versus field effect mobility $\mu$. The hBN-supported devices have mobility ranging from 20,000 to 100,000 cm$^2$/Vs, and the $\nu=-2$ state remains unresolved at $B$=8 T in all but one device. In comparison, the suspended devices have mobility ranging from 4,000 to 200,000 cm$^2$/Vs. Despite the two orders of magnitude variation in mobility, the $\nu=-2$ state in all suspended devices is resolved at or below 0.5 T, including ones with relatively low mobilities. The much smaller $B_{min}$ in suspended samples indicates a significantly larger $\Delta_{\nu=-2}$ than their hBN-supported counterparts.

Since $\Delta_{\nu=-2} \approx |\gamma_2/2 + 3\delta_2|$ corresponds to the gap in the BLG-like branch, our results demonstrate that, at the very least, the introduction of hBN substrates significantly modifies the hopping parameters $\gamma_2$ and $\delta_2$. In fact, by quantifying the hopping parameters, we find that the impact of the hBN substrate is more dramatic than simply shifting the bands. Here we fit the experimentally obtained crossing points to the LL spectra calculated by a $k \cdot p$ continuum Hamiltonian[23]. In addition to the crossing points in $R_{xx}(n,B)$ or $G(n,B)$ data, we also consider crossings at a constant magnetic field as a function of $n$ and out-of-plane electric field $E_\perp$. Fig. 4a displays such a data set for hBN1 at $B$=8 T, where the vertical blue stripes represent incompressible states at integer filling factors with $R_{xx} \sim 0$, interrupt by LL crossings points with high resistance (white or pink) peaks. As $E_\perp$ varies, crossings are observed at all integer plateaus at $\nu>-1$. The observation of crossings at $\nu=-2$ at relatively small $E_\perp \sim \pm 55$ meV is particularly informative: it arises from the crossing of the valley-split lowest LLs of the BLG-like bands. As $E_\perp$ depresses the K' valley while leaves K valley levels unchanged[27, 47], the observed crossing and the small corresponding small $|E_\perp|$ indicate that, at $E_\perp$=0, the (B, K',0) level has higher energy than (B, K, 1)[38]. This is also observed in a prior study[33]. The calculated low energy LL spectrum for device hBN1 is shown in Fig. 4b.

In suspended samples, however, in order to simultaneously account for a large $\Delta_{\nu=-2}$ and a high $B_P$ for crossing point P, we have to place the (B, K, 0/1) LLs energetically above the (B, K', 0/1) LLs. This is further verified by the absence of crossing in $\nu=-2$ at small $E_\perp$ at $B$=5.5 T for a suspended device S2 (Fig. 4c). Thus the proximity of hBN substrates not only modifies the hopping parameters, but also causes a "valley inversion" in the lowest Landau level of the BLG-like branch, as shown by the calculated LL spectrum for device S1 (Fig. 4d).

The hopping parameters of both devices, obtained from fitting the data, are summarized in Table 1. The most significantly modified parameters are $\gamma_2$ and $\delta_2$, which, after the introduction of hBN substrates, change by -280% and 560%, respectively. The impact of the hBN substrate on the band structure is qualitatively understood by a DFT-based simulation. Fig. 4e illustrates the calculated charge redistribution upon the addition of the hBN layers, where the blue (red) isosurface represents the region with net charges decrease (increase). We find that partial charges from TLG move toward hBN, and the wave functions of the top and bottom layers of TLG shift away from each other. Since $\gamma_2$ is the coupling between two outer layers of TLG, its magnitude is suppressed with the addition of hBN; $\delta_2$, which is the difference in energy between the middle layer and the average of the outer layers, increases accordingly.

Taken together, our observation indicates that substrates, or the lack thereof, have much larger impact on the band structure and the LL spectrum than previously thought possible. For graphene supported on hBN substrates, the twist angle between these two lattices systematically modify the band structure, even at very large angles. Thus the band structure of a given 2D material is tunable by substrates, and by twist angles. Such tunability should be taken into account when interpreting electronic and optoelectronic studies of 2D materials, and could be employed to tailor the band structures of 2D materials to optimize their applications.

**Acknowledgement**
The experiments are supported by DOE BES Division under grant no. ER 46940-DE-SC0010597. The theoretical works and collaboration between theory and experiment is enabled by SHINES, which is an Energy Frontier Research Center funded by DOE BES under Award # SC0012670. Growth of hBN crystals was supported by the Elemental Strategy Initiative conducted by the MEXT, Japan and a Grant-in-Aid for Scientific Research on Innovative Areas "Science of Atomic Layers" from JSPS.

Fig. 1. (a). Typical LL spectra of TLG. Red (blue) lines are LLs from MLG- (BLG-) like band, respectively. Solid (dashed) lines address K (K') valley, respectively. S, P and Q indicate LL crossing points. (b). False color SEM image of a dual-gated suspended TLG device. Scale bar: 1 μm. (c). Schematic of the dual-gated hBN-supported TLG device. (d). SEM image of a TLG (dark) transferred onto hBN (bright), the red (blue) dashed line indicates the long, straight edge of TLG (hBN) used to characterize the twist angle $\theta$.

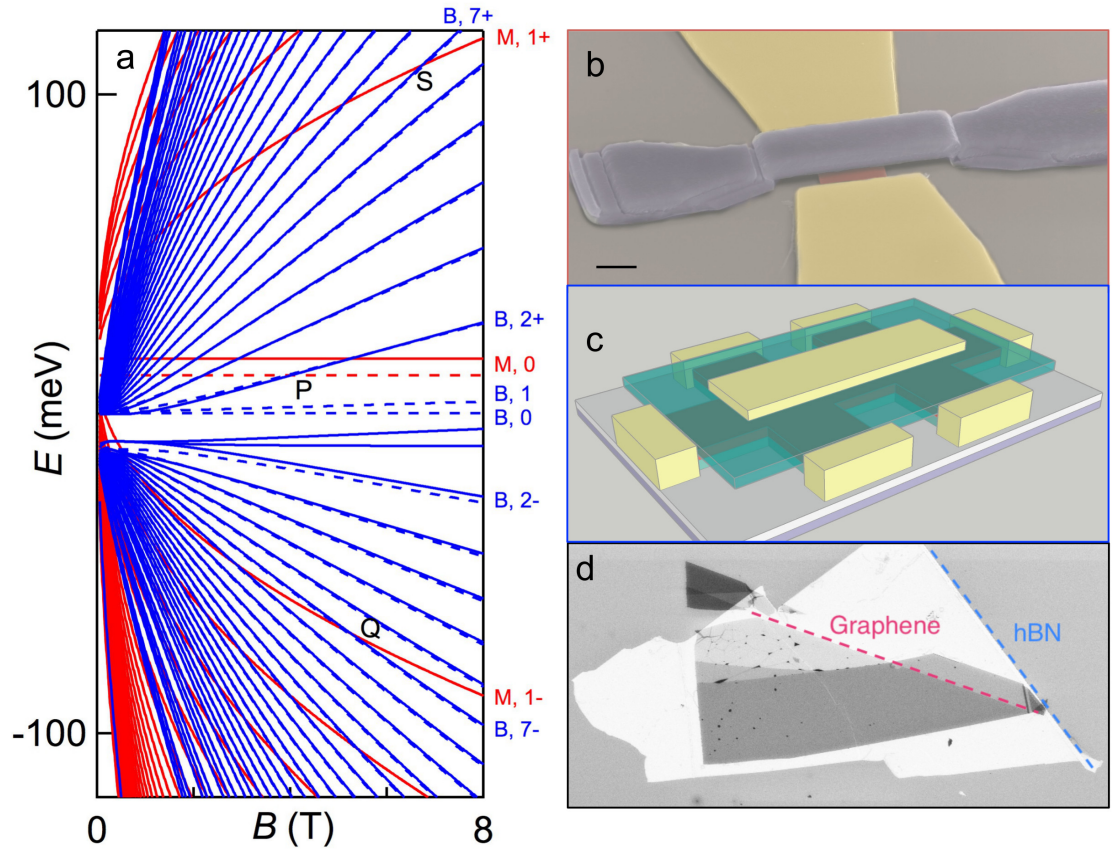

Fig. 2. (a-b). Landau fan $R_{xx}(n, B)$ at $E_\perp=0$ for devices hBN1 and hBN2, respectively. The unit is kΩ. (c-d). Low charge density zoom-in plot of Figure 2a and 2b, respectively. (e). $dG/dB$ $(n, B)$ fan diagram of suspended device S1.

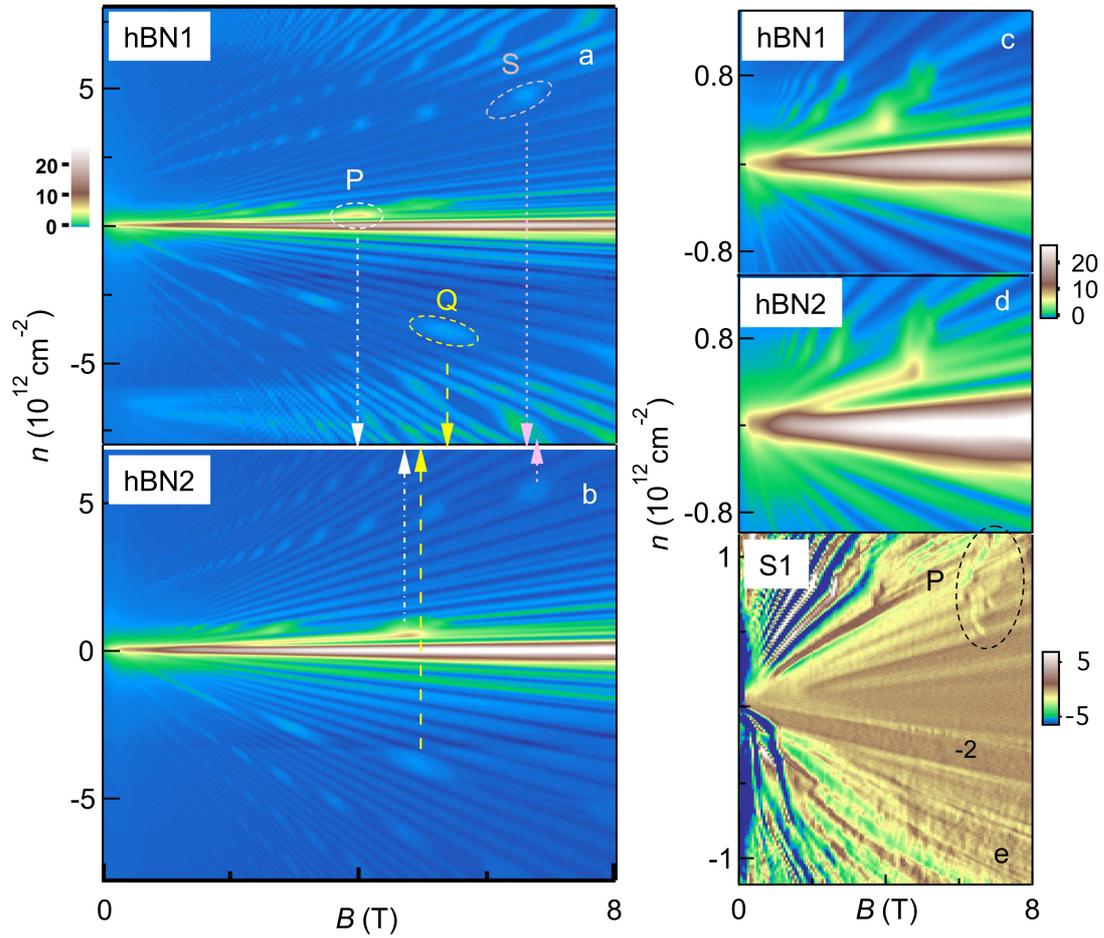

Fig. 3. (a-c). LL crossing points $B_P$, $B_Q$, $B_S$ as a function of twist angle $\theta$ between TLG and hBN substrate, respectively. Right axis of (a) plots the extracted hopping parameter $\gamma_2$ vs $\theta$. (d). The minimum magnetic field $B_{min}$ at which QH state $\nu=-2$ is resolved versus field effect mobility $\mu$. The markers of triangular (squares) denote suspended (hBN-supported) devices, respectively.

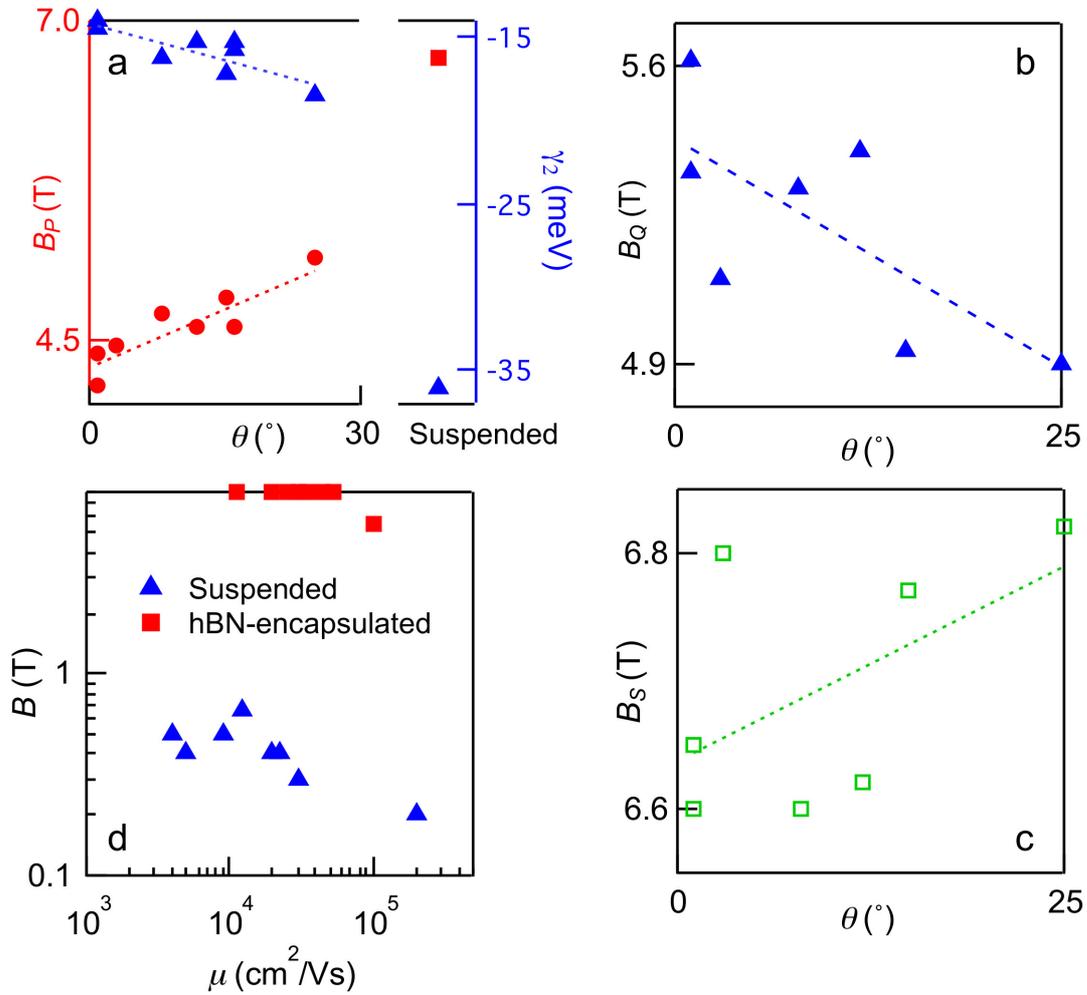

Fig. 4. (a). $R_{xx}(v, E_\perp)$ map at $B$=8 T of device hBN1. The unit is kΩ. (b). $G(v, E_\perp)$ map at $B$=5.5 T for device S2. The unit is $e^2/h$. (c-d). Low energy LL spectrum simulated for device hBN1 and S1 respectively. Red (blue) lines are LLs from MLG- (BLG-) like band. Solid (dashed) lines address K (K') valley, respectively. The expressions indicate the energies of the lowest LLs in terms of hopping parameters. (e). DFT simulated charge redistribution in hBN/TLG heterostructure. TLG layers are in brown, hBN layers are in white/green, the red (blue) isosurfaces denote the charger increase (decrease) regions when adding on hBN layers.

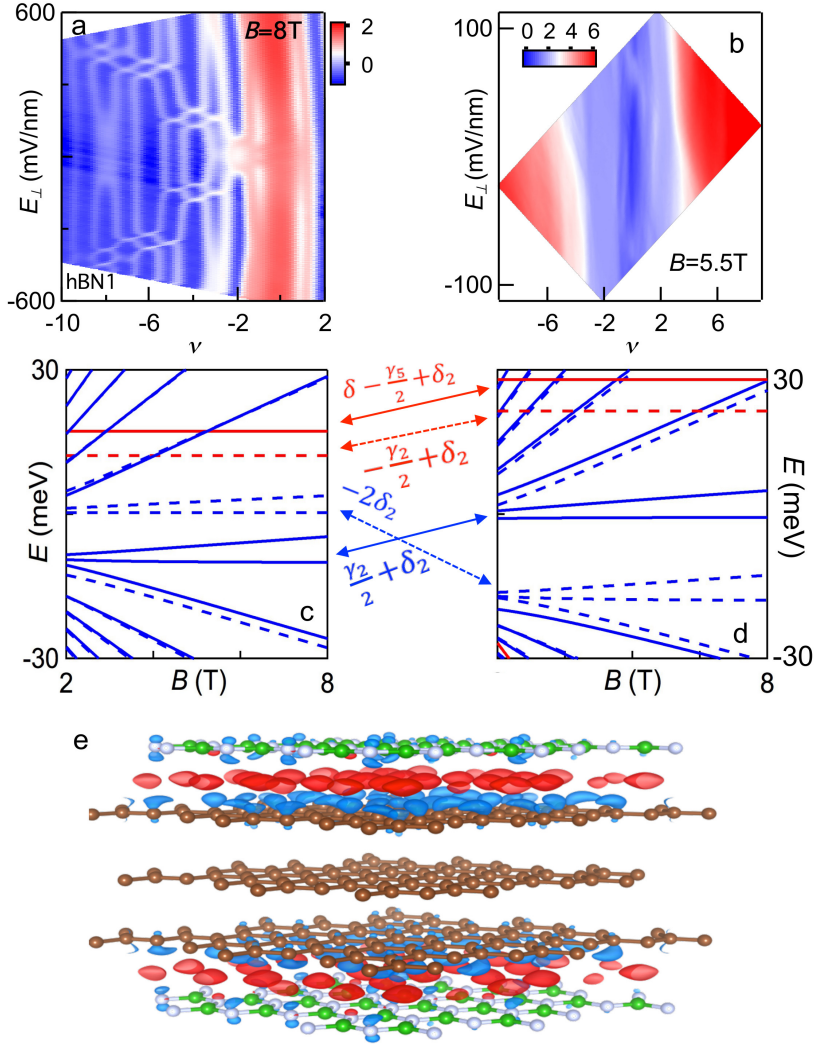

Table 1. Hopping parameters extracted from experimental data for a suspended and an hBN-supported TLG device.

| Device | $\gamma_0$ (meV) | $\gamma_1$ (meV) | $\gamma_2$ (meV) | $\gamma_3$ (meV) | $\gamma_4$ (meV) | $\gamma_5$ (meV) | $\delta$ (meV) | $\delta_2$ (meV) |
|---|---|---|---|---|---|---|---|---|
| S1 | 3100 | 355 | −41 | 315 | 150 | 40 | 47 | 1 |
| hBN1 | 3100 | 355 | −12.8 | 315 | 150 | 40 | 31.5 | 5.8 |